\documentclass[twocolumn,showpacs,preprintnumbers,amsmath,amssymb,10pt,prl]{revtex4}

\usepackage{graphicx}% Include figure files
\usepackage{dcolumn}% Align table columns on decimal point
\usepackage{bm}% bold math
\usepackage{hyperref}

\begin{document}
\title{Probing Loop Quantum Gravity with Evaporating Black Holes}
%\author{Sex, Truck \& Rock'n Roll}

\author{A. Barrau}%
 \email{aurelien.barrau@cern.ch}
\affiliation{%
Laboratoire de Physique Subatomique et de Cosmologie, UJF, INPG, CNRS, IN2P3\\
53, avenue des Martyrs, 38026 Grenoble cedex, France
}%

\author{X. Cao}%
 \email{xiangyu.cao@polytechnique.edu}
\affiliation{%
Ecole Polytechnique\\
Route de Saclay, 91128 Palaiseau, France
}%

\author{J. Diaz-Polo}%
 \email{jacobo@phys.lsu.edu}
\affiliation{%
Institute for Gravitation and the Cosmos \& Physics Department\\
Penn State, University Park, PA 16802-6300, USA\\
and\\
Department of Physics and Astronomy\\
Louisiana State University, Baton Rouge, LA 70803-4001, USA
}%

\author{J. Grain}
 \email{grain@apc.univ-paris7.fr}
 \affiliation{%
Institut d'Astrophysique Spatiale, Université Paris-Sud, CNRS\\
B\^atiments 120-121, 91405 Orsay cedex, France
}

\author{T. Cailleteau}%
 \email{thomas.cailleteau@lpsc.in2p3.fr}
\affiliation{%
Laboratoire de Physique Subatomique et de Cosmologie, UJF, INPG, CNRS, IN2P3\\
53, avenue des Martyrs, 38026 Grenoble cedex, France
}

\date{\today}

\begin{abstract}

This letter aims at showing that the observation of evaporating black holes should
allow the usual Hawking behavior to be distinguished from Loop Quantum Gravity (LQG) expectations. 
We present a full Monte Carlo simulation of the evaporation in LQG and statistical tests that
discriminate between competing models. We conclude that contrarily to what was commonly thought,
the discreteness of the area in LQG leads to characteristic features that qualify evaporating black
holes as objects that could reveal quantum gravity footprints.

\end{abstract}

%Use showkeys class option if keyword

\maketitle

\paragraph{Introduction--}
Loop Quantum Gravity (LQG) is a promising framework to nonperturbatively quantize General Relativity 
(GR) in a background invariant way (see \cite{lqg_review} for introductory reviews). Interestingly, 
it has now been demonstrated that different
approaches, based either on  quantizations (covariant or canonical) of GR, or on 
a formal quantization of geometry lead to
the very same LQG theory. As for any
tentative theory of quantum gravity, experimental tests are however still missing.
Trying to find possible observational signatures is obviously a key
challenge. In this article we address the following question : could there be objects in the contemporary
universe whose observation would lead to a clear signature of LQG ? Fortunately, the answer turns out 
to be positive.
Although small black holes have not yet been directly observed, they could have been formed by different
mechanisms in the early universe (see, {\it e.g.}, \cite{carr} for a recent review) or even by 
particle collisions. We don't review 
here the well-known possible production mechanisms, but instead we focus on how to use the evaporation of 
microscopic black holes to investigate the discriminating power of the 
emitted spectrum. Three different possible signatures will be suggested. 
Although one should be careful when pushing the limits of the LQG approach to black holes to 
the microscopic limit, our results rely on features of the area spectrum and are rather 
insensitive to small modifications in the theoretical framework.

\paragraph{Theoretical Framework--}
The state counting for black holes in LQG relies on the isolated horizon framework (see, {\it
e.g.},
\cite{diaz1} for an up-to-date detailed review). The isolated horizon is introduced as a boundary of the
underlying manifold before quantization. For a given area $A$ of a Schwarzschild black hole horizon, the
physical states arise from a punctured sphere whose punctures carry quantum labels (see, {\it e.g.},
\cite{diaz2} for a detailed analysis). Two labels $(j,m)$ are assigned to each puncture, $j$ being a
spin half-integer carrying information about the area and $m$ being its corresponding projection
carrying information about the curvature. They satisfy the conditions
\begin{equation}
\label{eq1}
A-\Delta\leq 8\pi \gamma \ell_P^2\sum_{p=1}^N{\sqrt{j_p(j_p+1)}}\leq A+\Delta,
\end{equation}
where $\gamma$ is the fundamental Barbero-Immirzi parameter of LQG, $\Delta$ is a ``smearing'' area 
and $p$ labels the different punctures, and
\begin{equation}
\label{eq2}
\sum_{p=1}^N{m_p=0},
\end{equation}
which corresponds to the requirement of a horizon  with spherical topology. Many specific features of the entropy
were derived in this framework \cite{diaz3}. Although the proportionality between the
entropy and the area is indeed recovered (when choosing correctly the $\gamma$ parameter) 
in the classical limit, the quantum structure still leaves a
clear footprint at microscopic scales.

Long ago, Bekenstein and Mukhanov postulated that due to quantum gravitational effects the area of a
black hole should be proportional to a fundamental area of the order of the Planck area \cite{bek} 
(the argument has recently been updated in \cite{gia}). 
This led to the idea of possible exciting probes of quantum gravity through associated lines in the
evaporation spectrum. However, following the pioneering work of Rovelli \cite{rovelli1}, it
was soon realized that the situation is drastically different in LQG where the spacing of the energy 
levels  decreases exponentially with the energy, therefore closing any hope for detection
\cite{rovelli2}. In (the first paper of) \cite {diaz3} a possible observational effect was 
suggested based on an exact computation of entropy and the observation of an effective 
discretization of it. In this letter we readdress this issue and show that at least three 
different
signatures can in fact be expected. Two of them are, as it could be expected, related with ``Planck scale''
black holes whereas the last one works also for larger black holes.\\
\begin{figure}[ht]
	\begin{center}
		\includegraphics[scale=0.45]{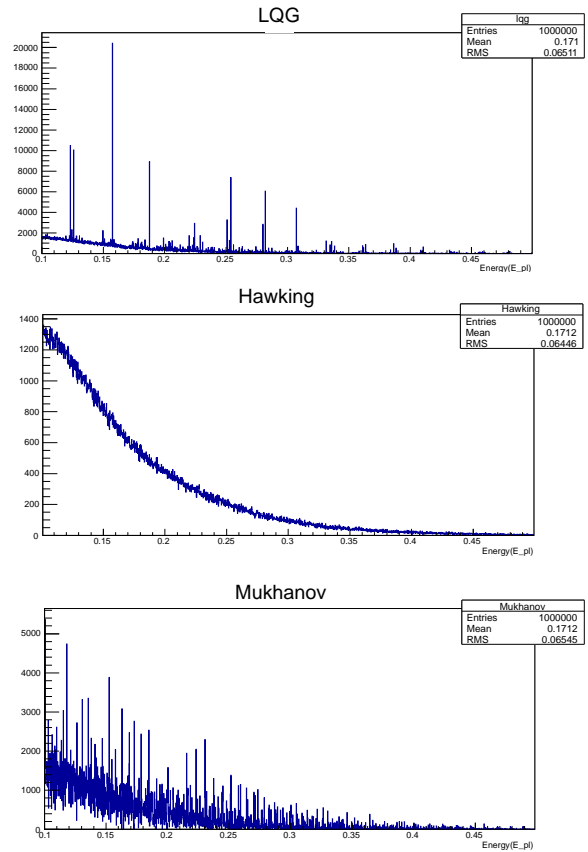}
		\caption{Spectrum of emitted particles in LQG, in the pure Hawking case, and in the
		Mukhanov-Bekenstein
		approach, from top to bottom.}
		\label{fig1}
	\end{center}
\end{figure}

\paragraph{Emission Lines in the Planck Regime--}

We first consider the evaporation of a black hole in the deep quantum regime. To this aim, we have
developed a dedicated and optimized algorithm. It is based on the ideas given in \cite{diaz1} 
and it was enhanced with an
efficient numeration scheme using a breadth-first search. As the projection constraint is very
time consuming, this improvement is mandatory to perform the computation up to high enough Planck 
areas. The evaporation is considered both according to the pure Hawking law and according to the LQG theory. In each case, we model the evaporation by
expressing the probability of transition as the exponential of the entropy difference multiplied by
the graybody factor. Arguments for the reliability and generality of this approach are given in
\cite{renaud}. As it can be seen from Fig.~\ref{fig1}, some specific lines associated with the
transitions occurring in the very last stages of the evaporation can be identified in the LQG
spectrum whereas the pure Hawking spectrum is naturally smooth. Two subtle points have to be taken
into account. First, the usually assumed ``optical limit'' of the graybody factors induces a heavy
distortion of the spectrum. The use of exact graybody factors, obtained by solving the quantum wave
equations in the curved background of the black hole, is in this case mandatory. To be maximally conservative,
we have used the very same graybody factors in the Hawking case and in the LQG case. Any
difference, as could be possibly expected due to an LQG-inspired metric modification 
(see, {\it e.g.} \cite{alesci}), 
would only make the discrimination between models easier. We have also assumed that the Hawking
evaporation stops at the  same mass as expected in LQG (namely $0.4~M_{Pl}$), once again to be
as conservative as possible. Second, even if one focuses on the ``high energy'' emission, say above
$0.15~M_{Pl}$, the contribution from states with a lower temperature is far from being negligible. We have
therefore pushed the computation of the area states, together with their multiplicity, up to $200~A_{Pl}$ to
ensure that the number of missed quanta remain smaller than a few percent.\\

\begin{figure}[ht]
	\begin{center}
		\includegraphics[scale=0.45]{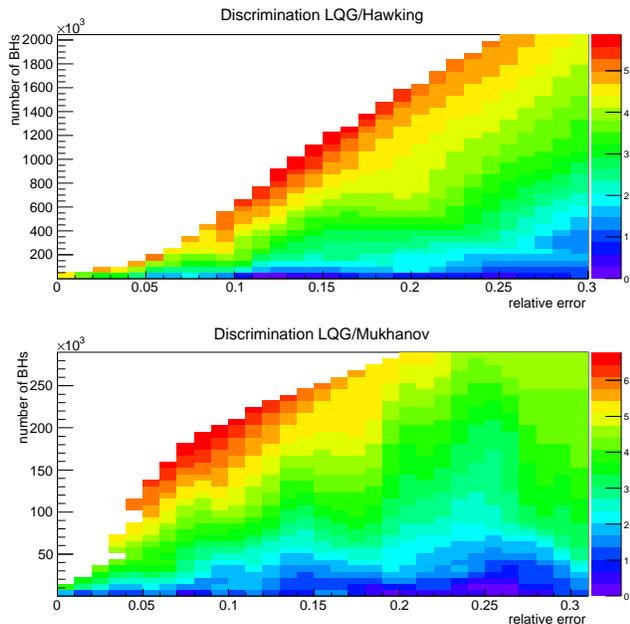}
		\caption{Number of evaporating black holes that have to be observed as a function
		of the relative error on the energy reconstruction of the emitted leptons for
		different confidence levels (the gray scale corresponds to the number of standard
		deviations). The first row corresponds to the discrimination between LQG and the
		Hawking hypothesis and the second row between LQG and 
		the Mukhanov-Bekenstein hypothesis.}
		\label{fig2}
	\end{center}
\end{figure}

Several Monte-Carlo simulations were carried out to estimate the circumstances under which the
discrimination between LQG and the standard behavior is possible. At each step, the energy of the
emitted particle is randomly obtained according to the relevant statistics and to the (spin-dependent)
graybody factor. Most simple statistical tests fail
to capture the intricate nature of the specific LQG features. We have therefore chosen to use a (slightly
improved) Kolmogorov-Smirnov (K-S) test. The K-S statistics quantifies the distance between
the cumulative distribution functions of the distributions and can be used for a systematic study of
possible discrimination (see, {\it e.g.} \cite{sho}. By investigating the K-S excess as a function of the energy, we have 
optimized the relevant interval for each relative error. As this latter is assumed to be known, it 
is meaningful to use it as an input for the statistical procedure. Figure~\ref{fig2} shows 
the number of black holes that should be
observed, for different confidence levels and as a function of the relative error on the
energy reconstruction, to discriminate the models. Clearly with either enough black holes or a relatively small error, a
discrimination is possible, therefore leading to a clear LQG signature. To still remain maximally
conservative, we have only considered emitted leptons. For a detector located nearby the black
hole, and due to the huge Lorentz factors, the electrons, muons and taus can be considered as
stable whereas quarks do not have enough time to fragment into hadrons.

For the sake of
completeness, we have finally implemented a K-S test between the LQG spectrum and the 
Bekenstein-Mukhanov one. Once again, the discrimination is possible with an even smaller
number of black holes as the lines are sitting at clearly different places.

Even if the Hawking and Mukhanov hypotheses are not expected to be reliable in the Planck era, this analysis shows
that a discrimination between LQG and other tentative approaches is possible.\\

\paragraph{Low-energy Emission in the Planck Regime--}

There is a second specific feature associated with the end point of the evaporation process. In LQG,
the last transitions take place at definite energies, of the order of the Planck  scale,
associated with the final lines of the mass
spectrum. On the other hand, in the usual Hawking picture, the most natural way to implement a 
minimal mass is to
assume a truncation of the standard spectrum ensuring energy conservation. Even if no minimal mass
is assumed, the spectrum has to be truncated to ensure that the black hole does not emit more
energy than it has. This is also the case in some string gravity models \cite{alex}. This leads to the important
consequence that the energy of the emitted quanta will progressively decrease and asymptotically
tend to zero. It is possible to distinguish this ``low-energy'' emission associated with the
end point from the (much more numerous) ``low-energy'' particles emitted before (when the black hole
temperature was lower) thanks to the dynamics of the process. For example, as soon as one considers
$\gamma$-rays with energies lower than $8\times 10^5$ GeV, the ``end point'' emission will take place at least
100~$\mu$s after the initial emission, making both signals easily distinguishable. Those ``relic''
quanta will be emitted with mean energies decreased by a factor 1/4 at each step (for
scalars and fermions). The time interval between consecutive emissions will typically 
increase with decreasing energies as $E^{-3}$. At 100 TeV, the mean interval is around 1~s.
This feature of the ``standard'' spectrum is therefore very different from the absence of 
low-energy
particles expected in the LQG case.

This probe should however be considered with care as it is less reliable than the two other
ones suggested in this work, being dependent on the specific assumption made for the
evaporation end point in the Hawking
case.

\begin{figure}[ht]
	\begin{center}
		\includegraphics[scale=0.45]{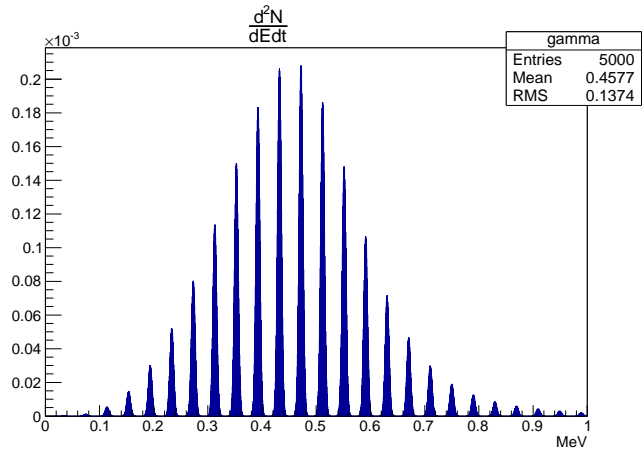}
		\caption{Instantaneous spectrum of a $\sim 100~{\rm keV}$ black hole taking into account the LQG
		modulation of the entropy.}
		\label{fig3}
	\end{center}
\end{figure}

\paragraph{Peaks in the Higher-Mass Regime--}
Up to now, the analysis was mostly focused on lines associated with the discreetness of the area, 
as could be seen on
Fig~\ref{fig1}. However, LQG specific features also lead to broader peaks in the spectrum, with a 
clear pseudoperiodicity, as shown in Fig~\ref{fig3}. 
Those peaks are associated with the ``large scale'' structure of the area spectrum. This periodicity has
been discussed in much detail (see \cite{diaz1} and references 
therein). We have observed this behavior up to 200 $A_{Pl}$ with an exact computation of the area
eigenvalues and we have checked it up to 400 $A_{Pl}$ with a dedicated Monte Carlo Markov Chain (MCMC) algorithm. 
Although some recent arguments suggest that this periodicity is damped for higher masses 
\cite{barbero2}, they cannot rule out the possibility of a ``revival'' of the periodicity
at larger areas (or even in the asymptotic limit), so it is relevant to study the possible 
observational effects that this periodicity would have in the macroscopic regime, in agreement
with the assumption made in most of the literature on the subject.
We here assume that it remains valid up to arbitrary large masses. This is not an unavoidable prediction of LQG
but this is clearly a possibility that arises, to the best of our knowledge, only in this framework. This makes
it a potentially interesting probe.
The mean area gap $d A$
between peaks can be shown to be independent of the scale. As, for a Schwarzschild black hole, $dA=32\pi
MdM$ and $T=1/(8\pi M)$, this straightforwardly means that $dM/T=cte$ where $dM$ refers to the mass gap
between peaks. This is the key point for detection~: in units of temperature, which is the natural energy
scale for the emitted quanta, the mass gap does {\it not} decrease for increasing masses. Any observable 
feature associated with this pseudoperiodicity can therefore be searched for through larger black holes.

This opens up the question of a possible detection of LQG effects with evaporating primordial black holes
(PBHs) in astrophysical circumstances. If PBHs were formed with a continuous mass spectrum $n_i(M_i)$,
where the subscript $i$ stands for initial values, it is now deformed according to $n(M)\propto M^2$
for $M<M_*$ and $n(M)\approx n_i(M)$ for $M>M_*$ where $M_*\approx 10^{15}$g is the initial mass of a
black hole whose lifetime is of the order of the age of the Universe. This is just due to the Hawking evaporation leading to ${\rm
d}M/{\rm d}t\propto M^{-2}$. In such a case, it is easy to show that the peak structure of the
instantaneous spectrum will be immediately washed out. The convolution of the individual spectra with the
mass distribution will lead to a Hawking-like $E^{-3}$ integrated spectrum. We have checked this expected
behavior with a Monte Carlo simulation. It should also be pointed out that the peak structure of the
``end-of-the-life'' spectrum, which is superimposed with the lines, is {\it not} due to the
pseudo periodic
structure of the entropy but to transitions to the last states, {\it i.e.} with the discreteness of the
area eigenvalues.

However, this does not at all close the issue of observing LQG features with astrophysical PBHs. The
continuous mass spectrum (typically scaling as $M^{-5/2}$) was a hypothesis
historically associated with a possible high normalization of the primordial power spectrum (or a very
blue tilt) which is ruled out by CMB observations.  Realistic models for PBH formation are now
associated with phase transitions (see, {\it e.g.}, \cite{kar}) or other phenomena leading to black holes
formed at a given mass $M_c$. If this mass is smaller than $M_*$, those black holes have already
disappeared. If $M_c>M_*$, that is if the horizon mass at the formation time was larger that $ 10^{15}$g,
those black holes are evaporating so slowly that their mass has nearly not changed. As not only  the
mass loss rate but also the area loss rate does decrease with the mass (${\rm d}A/{\rm
d}t\propto 1/M$), the peak structure exhibited in Fig.~\ref{fig3} should be observed from such black
holes. In this case, the instantaneous spectrum, together with its peak structure, can directly be probed. If the
mass is higher than typically $10^{17}$g the black hole will emit only massless particles, that is photons ($\sim 12\%$)
and neutrinos ($\sim 88\%$). The electromagnetic signal is not anymore contaminated by
$\gamma$-rays due to the decay of
neutral pions as quarks cannot be emitted. Although the redshift integration will slightly smear out the structures,
a very clean signature can therefore be expected as no mass integration is involved
developped in the possibly observed
signal. In addition, one can show that the total number of photons received per second by a detector of area $S$ can be written
as $\Phi\sim 10^4\times\frac{\rho_{PBH}}{\rho_c}\times\left(\frac{10^{17}~{\rm g}}{M}\right)^2\times S$ where $\rho_c=3H^2/8\pi G$
is the ``cosmological'' critical density and $\rho_{PBH}$ is the density of primordial black holes. This leads to a macroscopic signal
for quite a large range of masses and densities.

\paragraph{Conclusion--}

In this letter, we have shown that the specific features of the area of black holes in loop quantum gravity can 
lead to observational signatures. Although detecting evaporating black holes is in itself a challenge, we have
established that footprints of the underlying quantum gravity theory might indeed be observed in this way. This opens
a possible new window to probe LQG.

\paragraph{Aknoledgements--}

We would like to thank Adeline Choyer with whom this study was initiated. This work was
partially funded by the NSF Grants No. PHY0854743 and No. PHY0968871, the Eberly research 
funds of Penn State, and Spanish MICINN 
Grant No. ESP2007-66542-C04-01.

\end{document}